\documentclass{aa}
\usepackage{txfonts}
\usepackage{graphics}
\usepackage{times,amssymb}
\usepackage{epsfig}
\usepackage{graphicx}

\def\sax{{\it BeppoSAX}}

\makeatletter
\def\@cite#1#2{(#1\if@tempswa , #2\fi)}
\def\@citex[#1]#2{\if@filesw\immediate\write\@auxout{\string\citation{#2}}\fi
  \def\@citea{}\@cite{\@for\@citeb:=#2\do
    {\@citea\def\@citea{;\penalty\@m\  }\@ifundefined
       {b@\@citeb}{{\bf ?}\@warning
       {Citation `\@citeb' on page \thepage \space undefined}}%
\hbox{\csname b@\@citeb\endcsname}}}{#1}}
\makeatother  

\begin{document}

\title{Comparative study of the two large flares from SGR1900+14 
with the \sax\ Gamma-Ray Burst Monitor}

\author{C.~Guidorzi\inst{1}
\and F.~Frontera\inst{1,2}
\and E.~Montanari\inst{1,3}
\and M.~Feroci\inst{4}
\and L.~Amati\inst{2}
\and E.~Costa\inst{4}
\and M.~Orlandini\inst{2}
}

\institute{Dipartimento di Fisica, Universit\`a di Ferrara, via Paradiso
12, I-44100 Ferrara, Italy 
\and Istituto di Astrofisica Spaziale e Fisica Cosmica, CNR,
Sezione di Bologna, via Gobetti 101, I-40129 Bologna, Italy
\and ISA ``A. Venturi'', Modena, Italy
\and Istituto di Astrofisica Spaziale e Fisica Cosmica, CNR,
Sezione di Roma, via Fosso del Cavaliere, I-00133 Roma, Italy
}

\date{Received \date; Accepted \dots}

\abstract{
We report on spectral and temporal results of the 40--700 keV observations, obtained 
with the Gamma-Ray Burst Monitor (GRBM) on board \sax, of the two large flares
from the Soft Gamma-ray Repeater SGR~1900+14 occurred on August 27, 1998 and 
April 18, 2001. From their intensity, fluence 
and duration, the first one was classified as ``giant'' and  the second
as ``intermediate''.The spectral results have been obtained with an improved 
response function of the GRBM. We find that the two events have similar spectral 
properties, but different temporal properties. The major difference concerns
the time profiles of the light curves, whereas the lack of evidence in the 2001
flare for the erratic time variability found at high frequencies (10--1000 Hz)
in the 1998 flare could be ascribed to lower counting statistics.
We discuss these results in the light of the magnetar model proposed for SGR sources.

}

\titlerunning{Large flares from SGR1900+14 with \sax}
\authorrunning{Guidorzi et~al.}

\maketitle

%%%%%%%%%%%%%%%%%%%%%%%%%%%%%%%%%%%%%%%%%%%%%%%%%%%%%%%%%%
\section{Introduction}
\label{s:intro}
%%%%%%%%%%%%%%%%%%%%%%%%%%%%%%%%%%%%%%%%%%%%%%%%%%%%%%%%%%
Soft-Gamma Repeaters (SGRs, see Hurley 2000\nocite{Hurley00} and
Woods 2003\nocite{Woods03a} for reviews) are X--/gamma--ray transient 
sources that unpredictably undergo periods of bursting activity, separated by 
 sometimes long intervals (from years to decades) of quiescence.

To date, the SGR class includes four sources (SGR0525$-$66, SGR1627$-$41, SGR1806$-$20 
and SGR1900+14) plus two candidates, SGR1801$-$23 (only two bursts detected; 
Cline et~al. 2000)\nocite{Cline00} and SGR1808$-$20 (one SGR-like
burst observed; Lamb et al. 2003\nocite{Lamb03}).
All confirmed SGRs, on the basis of their early determined positions, appeared to be
associated with young supernova remnants (SNRs) of ages $\le 10^4$~yr. However, basing
on more precise locations, in most cases this association has been questioned 
\cite{Lorimer00,Hurley99a,Kaplan02a} and in some cases attributed to random
chance \cite{Gaensler01}. All SGRs appear to be
in our galaxy, except for SGR0525-66 which is in the Large Magellanic Cloud.

Typically, bursts from SGRs have short durations($\sim$0.1~s),
recurrence times of seconds to years, energies of $\sim10^{41} D^2_{10}$~ergs
($D = 10 D_{10}$~kpc). Their hard X--ray spectra ($>$25 keV) are analytically
consistent with an Optically Thin Thermal Bremsstrahlung ({\sc ottb}) with temperatures 
of 20--40 keV.
During quiescence, persistent X-ray emission ($<$ 10 keV) has been observed 
from all of them  with luminosities of $\sim 10^{35}$--$10^{36} D^2_{10}$~erg~s$^{-1}$ 
and power-law spectral shapes. In the case of SGR1900+14, an additional blackbody 
({\sc bb}) component ($kT_{bb} \sim 0.5$ keV) is requested \cite{Woods03a}.
SGR1806$-$20 and SGR1900+14, during quiescence, show X-ray pulsations with periods 
in the range from 5 to 8 s and spin-down rates of $\sim 10^{-11} - 10^{-10}$ s/s 
(see, e.g., Hurley et~al. 1999b, 
Kouveliotou et al. 1999\nocite{Hurley99b,Kouveliotou99}). From these sources, also
evidence of X--ray lines has been reported during bursts:
an emission line at $\sim 6.4$~keV from the former source \cite{Strohmayer00} and an
absorption--like feature at $\sim 5$~keV from the latter \cite{Ibrahim02}. 

Rarely, ``giant'' flares of hard X--/$\gamma$--rays have been observed
from SGR0525-66 \cite{Mazets79} and SGR1900+14 
\cite{Cline98,Feroci99,Hurley99a,Mazets99b}. They show durations of hundreds of 
seconds, pulsations during most part of the event,
and peak luminosities in excess of $10^{44} D^2_{10}$~erg/s. 

After large bursts fading afterglow emission has been observed.
A fading X--ray afterglow, visible for several days, has been discovered
after the large flares from SGR1900+14 (i.e., the giant flare of August 1998 and the 
intermediate flare of April 2001, see below). The X--ray afterglow  
decays as a power-law ($F(t)\propto t^{-\alpha}$) with temporal index 
$\alpha\sim0.7$ after the August 1998 event \cite{Woods01}
and $\alpha\sim0.9$ after the April 2001 event \cite{Feroci03}. The X-ray 
afterglow spectrum is the combination of a power-law ({\sc pl}) and a 
{\sc bb}, with the {\sc bb} not visible at early times when the {\sc pl} component 
is predominant, but emerging at later times, when the {\sc pl} component 
becomes weaker \cite{Woods99a,Feroci03}, 
suggesting that the non-thermal component fades more rapidly than the thermal one 
(see also Lenters et~al. 2003,\nocite{Lenters03} for the X-ray tail that followed the 
short burst occurred on April 28, 2001). The presence of  X--ray afterglow emission
from the other SGR sources is still an open question.
Radio afterglow has also been observed from SGR1900+14 after the August 1998 event 
by Frail et al. (1999)\nocite{Frail99}, who detected a transient radio source with
the {\it Very Large Array} telescope at the source position following the giant flare.
This is the only point-like radio source associated to an SGR to date.

On the basis of their locations and temporal properties and
the absence of companion stars, SGRs have been proposed to be young ($< 10^4$ yrs)
isolated neutron stars (NS) with ultra-strong magnetic fields
($B_{dipole} > 10^{14}$ gauss), a.k.a. ``magnetars''.
The magnetar model \cite{Duncan92,Thompson93,Thompson95} considers a young
neutron star with a very strong magnetic field ($\sim
10^{14} - 10^{15}$ G), whose decay powers the quiescent X-ray emission
through heating of stellar crust, while the low-level seismic activity
and the persistent magnetospheric currents \cite{Thompson02}
occasionally cause big crustquakes which trigger short bursts and
large flares. In the magnetar scenario, the absorption feature from SGR1806-20
can be interpreted as ion-cyclotron resonance in the huge magnetic field of the NS
\cite{Ibrahim02}.

SGRs share some properties (pulse period distribution, spin-down rate,
lack of a companion star, quiescent X-ray luminosity) with a 
peculiar class of neutron stars, the so--called anomalous X-ray pulsars
(AXPs, see, e.g., Mereghetti 1999\nocite{Mereghetti99} for a review).
Additional evidence for a link between the two classes
has been provided by the detection of SGR-like bursting activity also from the  
AXPs 1E~2259+586 (Kaspi et al. 2002, 2003)\nocite{Kaspi02,Kaspi03}
and 1E1048.1-5937 (Gavriil et al. 2002),\nocite{Gavriil02} and from the recent 
discovery of an absorption--like feature at $\sim 8.1$~keV from the 
AXP 1RXS~J170849$-$400910 \cite{Rea03}.

\subsection{SGR1900+14}

SGR1900+14 was discovered in 1979 following three bursts in two days
\cite{Mazets79}. After its discovery, the source was found bursting again in 1992
\cite{Kouveliotou93} and, after five years of quiescence, in May 1998, when
it entered an extremely active bursting period, that reached its maximum with
the above mentioned giant flare of August 27, 1998.
%\cite{Cline98,Feroci99,Hurley99a,Mazets99}. 
A precise localization of the source with IPN \cite{Hurley99c} showed that SGR1900+14 lays
just outside a Galactic SNR, G042.8+00.6 and could be associated with it.
However, recently Kaplan et al. (2002b)\nocite{Kaplan02b} found three new candidate
SNRs (G043.5+00.6, G042.0-00.1 and G041.5+00.4), that could be related to
SGR1900+14 as well.
Observations of the quiescent soft X--ray counterpart \cite{Vasisht94}
have shown a 5.16-s periodicity with a spin-down rate
of $10^{-10}$ s/s \cite{Hurley99b,Kouveliotou99}.

Several measurements of the quiescent spectrum have been performed (e.g., Hurley et al. 
1999b, Kouveliotou et al. 1999).\nocite{Hurley99b,Kouveliotou99}
An X--ray observation with the \sax\ satellite \cite{Woods99a} shows that
the 0.1-10~keV quiescent spectrum can be described by a photoelectrically 
absorbed  ($N_{\rm H} \sim 1.8 \times 10^{22}$~cm$^{-2}$) {\sc bb}
($kT_{bb} \sim 0.5$ keV) plus a {\sc pl} with a photon index $\Gamma = 1.1 \pm 0.2$.

The source spectrum during the standard bursting activity has been mainly observed in the 
hard ($>20$ keV) X--/gamma--ray band. Results reported by Mazets et al. (1999a)
from {\it Konus--Wind} observations show that the burst photon spectra can be 
analytically described by an {\sc ottb} model ($I(E)\propto E^{-1} \exp{(-E/kT)}$) 
with $kT \sim 20-30$~keV with no significant spectral
evolution within a single event or from event to event. Only in the case of a few bursts, 
discussed by Woods et al. (1999b)\nocite{Woods99b}, the spectrum is better described
by the smoothly broken power-law, widely used to describe GRB spectra \cite{Band93}.
In these cases also a soft-to-hard spectral evolution has been observed,
with hardness/intensity anticorrelation.

The giant flare occurred on 1998 August 27 (here after GF98) was observed with  the 
Konus--{\it Wind} spectrometer \cite{Cline98,Mazets99b}, the {\it Ulysses} burst monitor 
\cite{Hurley99a} and the \sax\ GRBM \cite{Feroci99}. The 5.16-s periodicity, along 
with its harmonics (e.g., Feroci et al. 1999 and 2001,\nocite{Feroci99,Feroci01}
hereafter F99 and F01, respectively), was clearly detected during the 
flare. Mazets et al. (1999b) find the spectrum  well described with 
an {\sc ottb} model with temperature $kT$ evolving rapidly 
(in about 1 s and in a non-monotonic way) from $>300$~keV to  $\sim 20$~keV.
Feroci et al. (1999) find a more complex 40--700 keV spectrum ({\sc ottb}, with 
$kT\sim 31$~keV, plus {\sc pl} with photon index $\Gamma \sim 1.5$) in the early part 
(first 68 s) of the event which evolves to an {\sc ottb}-like shape ($kT \sim 29$~keV) at late 
times (last 128 s). At that time, as pointed out by F99, 
the response function of the \sax\ GRBM was not well known at large instrumental 
off--axis angles, in the direction of which the source was observed, and thus a 
systematic error of 10\% was tentatively added to the statistical uncertainties.
By joining together the \sax\ GRBM spectra with the 20--150 keV {\it Ulysses} data (F01), 
also affected by similar systematics, the first 128-s spectrum  after the 68 s 
from the flare onset was fit by two {\sc bb}, with $kT_{\rm bb1} \sim$ 9~keV and 
$kT_{\rm bb2} \sim$ 20 keV, plus a {\sc pl} model with photon index $\Gamma \sim$~2.8,
while the later 128-s spectrum was better described by an {\sc ottb} model with 
$kT \sim 29$~keV, consistent with the F99 results.

After the giant flare of 1998 August 27 and the recurrent bursting activity
prolonged until 1999 \cite{Mazets99a}, the source entered a period of quiescence
for more than two years, which ended on 2001 April 18, when another large flare 
(hereafter IF01) with a shorter duration and  intermediate intensity occurred. Due 
to a simultaneous proton solar burst, both the {\it Ulysses} burst monitor and the 
Konus--{\it Wind} spectrometer were overwhelmed by a high count rate. The only 
instrument which provided high-time resolution data of the event was the \sax\ GRBM
(Guidorzi et~al. 2001a; 2001b)\nocite{Guidorzi01a,Guidorzi01b}.

An improved response function of the \sax\ GRBM is now available
(see below) for all off--axis angles.  
In this paper we present the results of the spectral analysis of the 2001 flare, 
the reanalysis of the spectral data of the 1998 giant flare with the new GRBM 
response function, and the results of the Power Spectral Density (PSD) estimate
of the high-frequency (up to 1 kHz) flux variations of the source during
the two flares. 
Preliminary results of this analysis were reported elsewhere \cite{Guidorzi03a}.

%%%%%%%%%%%%%%%%%%%%%%%%%%%%%%%%%%%%%%%%%%%%%%%%%%%%%%%%%%
\section{The \sax\ GRBM}
%%%%%%%%%%%%%%%%%%%%%%%%%%%%%%%%%%%%%%%%%%%%%%%%%%%%%%%%%%
\label{s:obs}

The Gamma-Ray Burst Monitor (GRBM; Frontera et~al. 1997; Feroci et~al. 1997)
\nocite{Frontera97,Feroci97} is one of the instruments on board the \sax\ satellite 
\cite{Boella97} operative through June 1996 and April 30, 2002. The GRBM
consisted of four optically independent CsI(Na) units forming
a square box: each unit had a geometric area of $\sim$ 1136 cm$^{2}$.
GRBM units No. 1 and 3  were co-aligned with the Wide Field
Cameras (WFC's; Jager et~al. 1997). \nocite{Jager97}
The data continuously available from the GRBM included
1-s ratemeters in the 40--700 keV  and $>$100 keV energy channels, 225--channel
spectra in the 40--700~keV band integrated over 128 s and, in the case of burst trigger,
7.8125-ms ratemeters for 96~s and, for 10~s after the trigger, $\sim 0.5$-ms count 
rates, both in the 40--700 keV band.
From the 1-s ratemeters, it is possible to extract the source 40--100 keV and
100--700 keV count rates under the assumption that the source flux above 700~keV
is negligible (see Amati et al. 1999\nocite{Amati99}).

The detection units, except for the units 1 and 3 within small off--axis directions
\cite{Amati99}, were surrounded, in  a complex way, by materials of the \sax\ 
payload. 
In order to exploit the capabilities of the GRBM units 2 and 4 and those of units 1 
and 3 at large off--axis angles, the complete GRBM response function was derived
with Monte Carlo methods (Rapisarda et~al. 1997, Calura et al. 
2000)\nocite{Rapisarda97,Calura00}.  Recently the complete response has been tested
with the Crab Nebula and with several GRBs which were
observed, at different instrument off--axis angles, with both the GRBM and the 
BATSE experiment (e.g., Guidorzi 2002, Guidorzi et al. 2003b, Costa \& Frontera 
2003)\nocite{Guidorzi02,Guidorzi03b,Costa03}, with the limitation of a generally
smaller statistical accuracy of the GRBM data, with respect to the BATSE data.
The deconvolution results of the GRBs, in terms of derived GRB direction and 
photon spectrum, are in excellent agreement with those obtained with BATSE.
Thus we expect that systematic errors, also for possible terrestrial albedo effects, 
are similar to those which affect the deconvolution of the BATSE data (e.g., Paciesas
et al. 1999\nocite{Paciesas99}).
Actually, we must point out that the total counts measured for the brightest GRBs
detected with both GRBM and BATSE and used to test the GRBM response function, amount to
a few $10^5$~cts in the most illuminated GRBM unit (e.g., $\sim250,000$ for GRB990123,
$\sim140,000$ for GRB991216, $\sim130,000$ for GRB971110),
while about $8\times 10^5$~cts were collected in the GRBM unit 1 due to GF98.
Due to the smallest statistical uncertainties in the latter case, the systematics
could have a major influence on deconvolution results.

%%%%%%%%%%%%%%%%%%%%%%%%%%%%%%%%%%%%%%%%%%%%%%%%%%%%%%%%%%
\section{Observations and data analysis}
%%%%%%%%%%%%%%%%%%%%%%%%%%%%%%%%%%%%%%%%%%%%%%%%%%%%%%%%%%

GF98 triggered the GRBM on August 27, 1998 at 10:22:15.7 UT, while
IF01 triggered the GRBM on April 18, 2001 at 07:55:11.5 UT.
IF01 occurred when the SGR1900+14 line of sight  was only $\sim 11^{\circ}$ 
off--axis from the detection unit No. 1 (indeed the flare was initially observed also 
with the WFC No. 1; Guidorzi et~al. 2001a, 2001b, Feroci et al. 2003).
\nocite{Guidorzi01a,Guidorzi01b,Feroci03}
GF98 occurred when the line of sight to the source
was at high off--axis angles: an elevation angle of 48$^\circ$ with 
respect to the GRBM equatorial plane, an azimuthal angle of 29$^\circ$ with respect to 
the GRBM unit 1 axis, and 61$^\circ$ with respect to unit 4 axis. % (F99)\nocite{Feroci99}.
All GRBM units detected the event, with the best signal given 
by  unit 1. Our results, except when expressly stated, will be based on the data 
obtained from this unit. 

%
% Figure 1
%
\begin{figure}[!t]
\resizebox{\hsize}{!}{\includegraphics{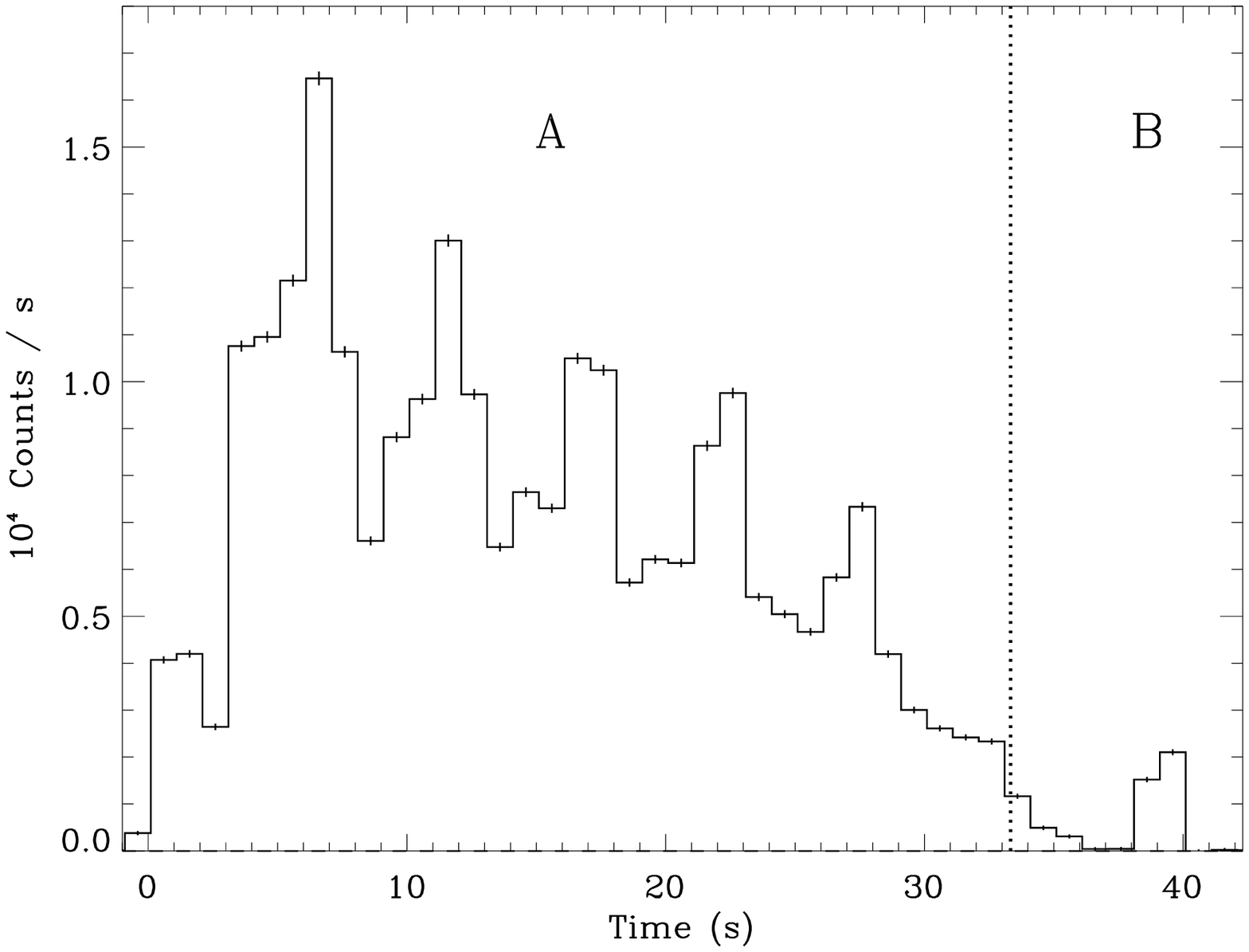}}
\resizebox{\hsize}{!}{\includegraphics{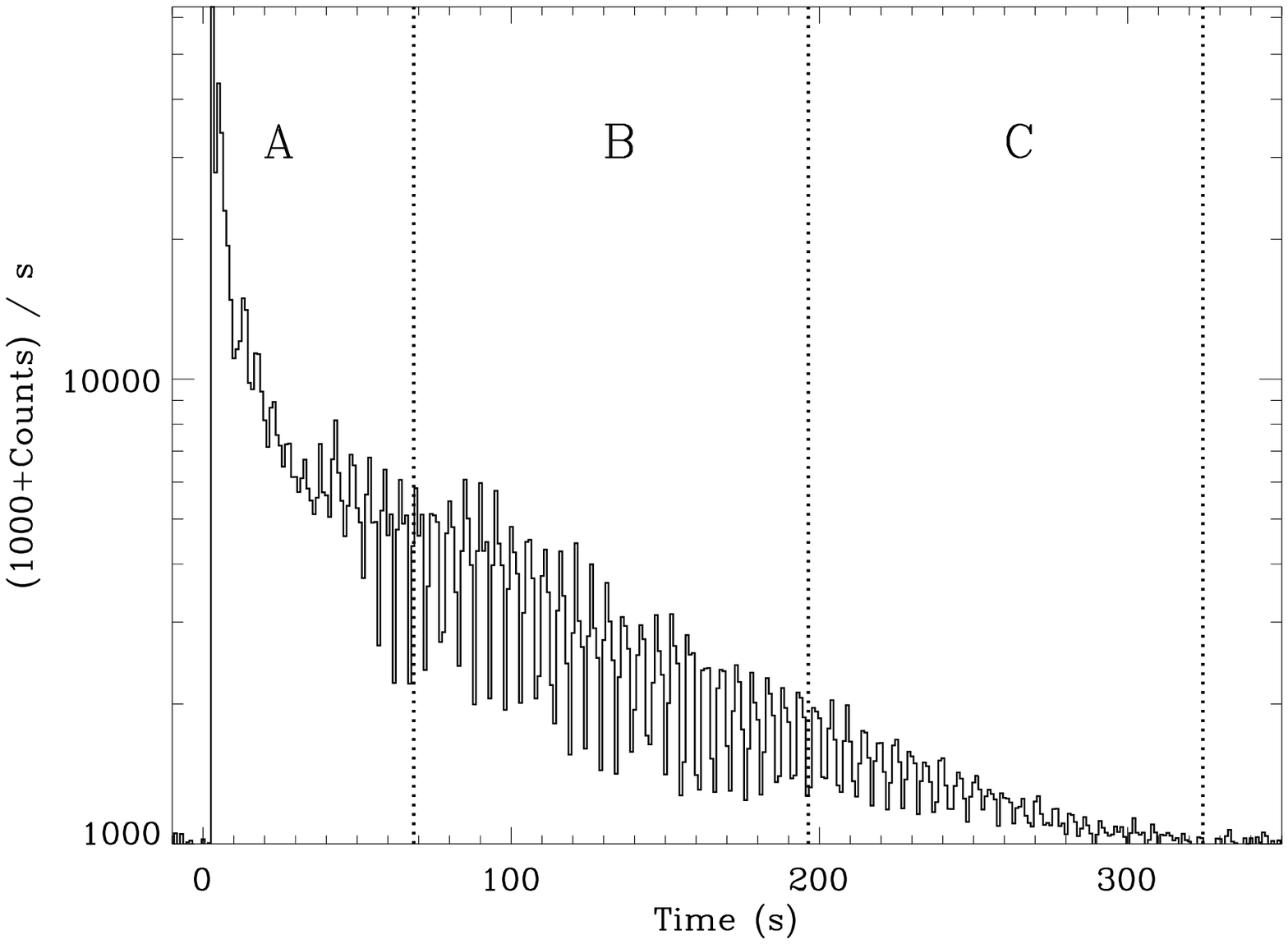}}
\caption[]{1-s background subtracted light curves of the two flares
 in the 40--700~keV energy band. {\it Top}: 2001 April 18 event;
{\it bottom}: 1998 August 27 event.
The $5.2$-s modulation is apparent in both cases. Vertical dotted lines
mark the 128-s intervals during which  225-channel spectra were accumulated on board.}
\label{both_flares_1900}
\end{figure}

The background subtracted light curves of both flares are shown in 
Fig.~\ref{both_flares_1900}, where the onset time is used as zero time.
Details of the observation of the GF98 event can be found in the paper by F01.
\nocite{Feroci99} The time duration of IF01 was 40~s,
whereas GF98 lasted about 300~s. The 7.8125-ms ratemeters cover the entire IF01.
The 0.5-ms ratemeters are available for the interval 0.69--10.69~s from the flare onset.
During IF01, the background level was fairly stable and was evaluated by 
linear interpolation of the background
levels in time intervals [-50, -10]~s and [+50,+150]~s, before and after the flare,
respectively. 

For the spectral analysis of IF01, the available data are two 225--channel 128-s
contiguous spectra. The start and end times of these spectra are reported in 
Table~\ref{t:log}. The interval A covers $\sim 97$\% of the 
flare fluence; the interval B includes little more than the  isolated pulse 
at $t\sim40$~s from the flare onset (see Fig.~\ref{both_flares_1900}).

The available data for the spectral analysis of GF98  are three spectra in three 
contiguous 128-s intervals  (A, B, C) described in Table~\ref{t:log}. 
The A spectrum includes both the initial hard spike and the intermediate smooth decay,
whereas the B and C spectra include the 5.16-s pulsation modulated decay 
(see Fig.~\ref{both_flares_1900}).

To get the source spectra, we subtracted from the above data a background spectrum of 
the same duration (128~s) interpolated between those measured before and after the 
flare.

The spectra were analyzed with the {\sc XSPEC} software package \cite{Arnaud96}.
The quoted errors are given at 90\% confidence
level (CL) for one parameter ($\Delta \chi^2 = 2.7$), except when otherwise
specified.

%
% Table 2
%
\begin{table*}
\begin{center}
\caption{Time intervals of the available 128-s count spectra.}
\begin{tabular}{cccc}
\hline
\noalign{\smallskip}
Flare &  Interval & Start time (s) & End time (s)\\
      &        &  from flare onset & from flare onset \\
\noalign{\smallskip}
\hline
\noalign{\smallskip}
2001 April 18 (IF01)	& A	& $-$94.67	& +33.33\\
			& B	& +33.33	& +161.33\\		  
\noalign{\smallskip}
1998 August 27 (GF98)   & A     & $-$59.6	& +68.4 \\
                        & B	& +68.4	  	& +196.4\\
			& C	& +196.4	& +324.4\\	
\noalign{\smallskip}
\hline
\noalign{\smallskip}

\end{tabular}
\label{t:log}
\end{center}
\end{table*}

\section{Results}

%------------------------------
\subsection{Light curves}
%------------------------------
The light curves of both flares are complex and different from each other.
As it can be seen from Fig.~\ref{both_flares_1900}, unlike GF98, IF01 does not 
shows any initial spike; the time duration of IF01 is much shorter than GF98,
ending after about 40~s when a four-peaked repetitive pattern
sets up in GF98 (F01). The light curves show measured counts, that do not account
for the different effective areas of the two flares: actually, the effective area
corresponding to the direction of GF98 was about 1/3 that of IF01.

% Figure 2
%
\begin{figure}
\resizebox{\hsize}{!}{\includegraphics{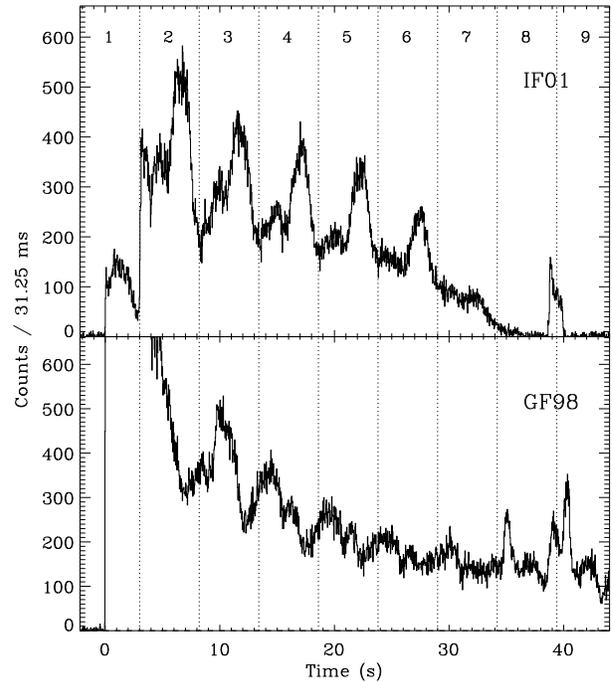}}
\caption[]{Background subtracted light curve  of IF01 compared with
the corresponding first 40~s of GF98 in the 40--700~keV energy 
band. {\it Top:} IF01; {\it bottom}: GF98. Vertical dotted lines mark the 5.2-s 
intervals and are synchronized with the IF01 pulsations.}
\label{both_40s_1900}
\end{figure}

\begin{figure}[!h]
\begin{center}
\resizebox{\hsize}{!}{\includegraphics{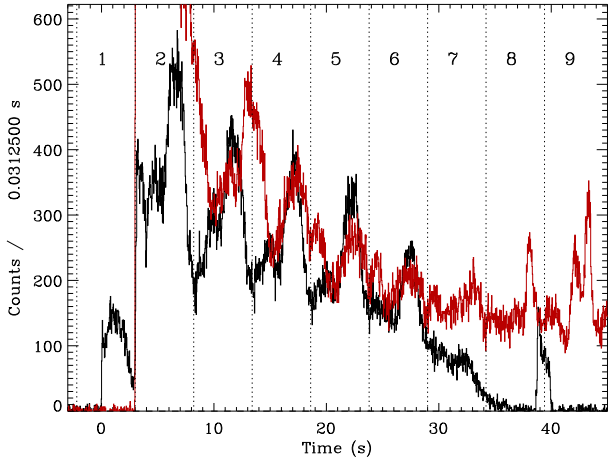}}
\caption[]{Light curves of the first $\sim$40~s of both IF01 ({\em black})
and GF98 ({\em red}) superposed to each other assuming that the two steepest rises
correspond to the onset of the flares.}
\label{both_offsets_1900_top}
\end{center}
\end{figure}

\begin{figure}[!h]
\begin{center}
\resizebox{\hsize}{!}{\includegraphics{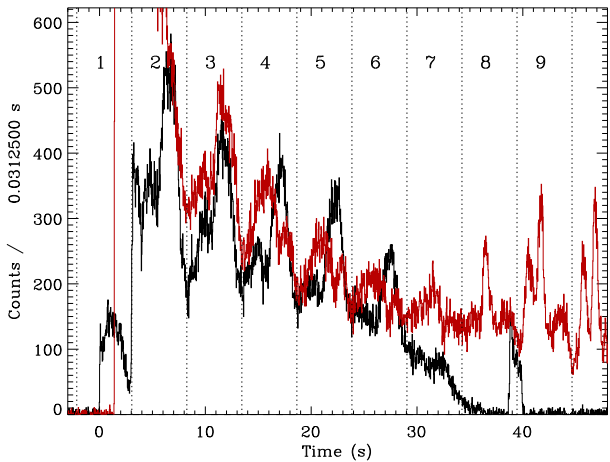}}
\caption[]{Light curves of the first $\sim$40~s of both IF01 ({\em black})
and GF98 ({\em red}) superposed to each other, with the first interpulses aligned.
The slices' boundaries have been chosen to mark the first minima.}
\label{both_offsets_1900_top_interpulse}
\end{center}
\end{figure}

In order  to better compare the time behaviour of IF01 with that of GF98 during
the early phase, in Fig.~\ref{both_40s_1900} we show the light curves of both 
events in the first 40 s.
Even if some similarities between the two light curves could be found, in the first 8 
seconds they are markedly different. Apart from the GF98 initial spike, which is not 
observed in IF01, in the first 3 seconds, IF01 
shows a first weak peak, which does not have a correspondence with any 
peak of the GF98 light curve. From the two light curves it 
is also unclear to which phase of the IF01 light curve the GF98 initial spike 
corresponds. We have considered two possible correspondence cases: i) the GF98 initial
spike time corresponds to the fast rise of the IF01 light curve 
(duration $\Delta\,t_{rise} = 0.25\pm0.05$~s) at 3~s from the flare onset 
(Fig.~\ref{both_offsets_1900_top}), ii) the dips corresponding to the interpulses 
in both light curves are aligned (see Fig.~\ref{both_offsets_1900_top_interpulse}). 
In the first case (see Fig.~\ref{both_offsets_1900_top}), we find that all IF01 dips 
soon after each peak are separated from the assumed main rise at multiple distances of
$\sim 5.2$~s. Using Fourier techniques, after detrending the data stretch by fitting
with a trapped-fireball model (see Fig.~\ref{f:fireball}), the best 
estimate of this periodicity in the time intervals from 2 to 7, is
$P_{01} = 5.21 \pm 0.05$~s (negligible barycenter correction), a value which is 
consistent with that measured from the observation of the X-ray quiescent source 
soon after the event ($P_{quiesc} = 5.17284268$~s) \cite{Woods03b}.
In GF98 the dips have period $P_{98}$ circa 5.16~s (F99),
but they are out of phase with respect to the initial spike.

In the second case (see Fig.~\ref{both_offsets_1900_top_interpulse}), the result is that
the GF98 initial spike corresponds to the peak of the weak pulse preceding the
first dip (and the main rise) of the IF01 light curve. 

% Figure 3
%

%
% Figure 4
%

In both cases, it is possible to see that, when the pulse peak fades away (slice 7
of Fig.~\ref{both_offsets_1900_top} or \ref{both_offsets_1900_top_interpulse}), the 
corresponding dip is no more visible. The pattern completely changes soon before IF01
ends (slice 8): while the 
continuum level seems to fade under the GRBM sensitivity, a pulse rises up to a peak 
count rate of about 160 counts in 31.25~ms, the same level as the early pulse. 
The pulse does not seem to have properties similar to the preceding
regular pulses: it neither occurs in phase with them nor exhibits a similar structure.

\begin{figure*}
\includegraphics[width=17cm]{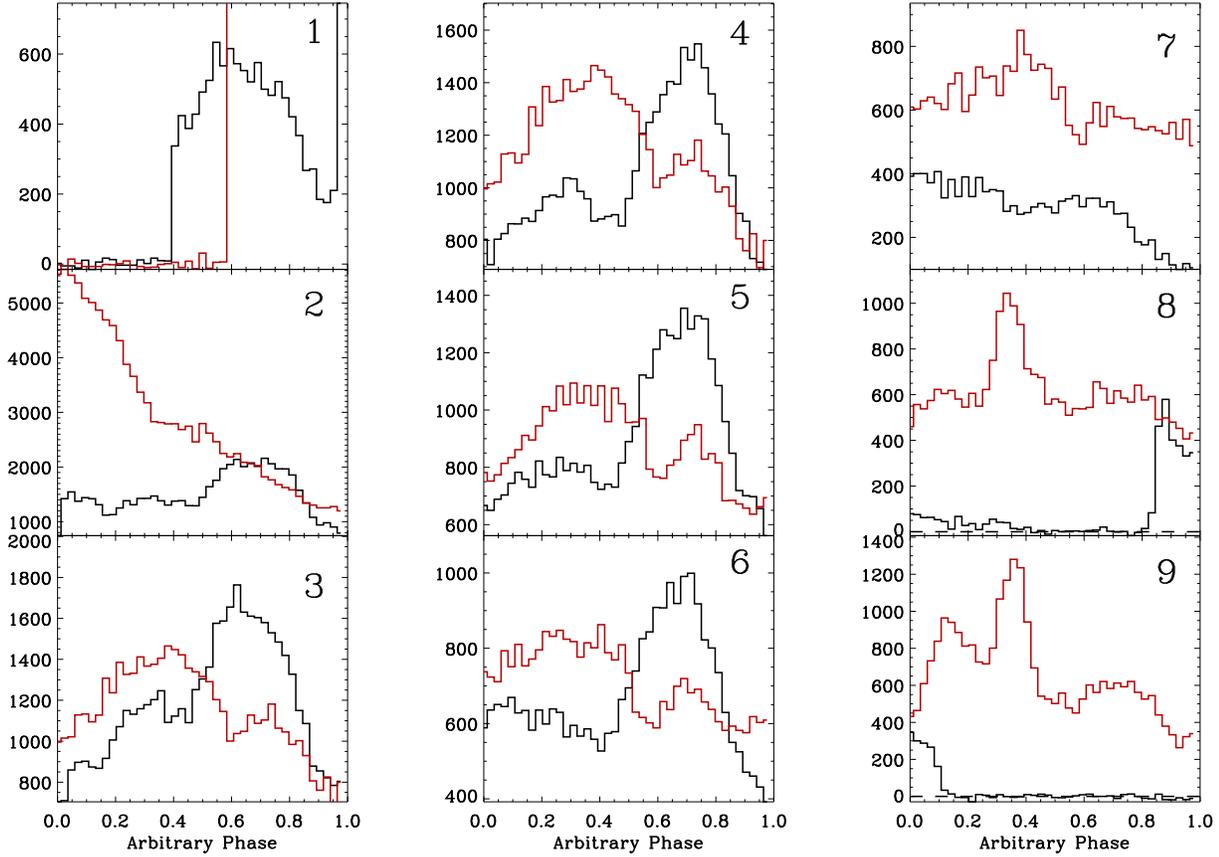}
\vspace{0.5cm}
\caption[]{Pulse shape evolution of both flares (IF01: {\em black}, GF98: {\em red}) during
the first nine slices. The time offset between the two events was chosen in order to
align the first interpulses (see also Fig.~\ref{both_offsets_1900_top_interpulse}).
{\em Horizontal scale}: phase of rotation cycle; {\em vertical scale}: counts per 125~ms.}
\label{pulseshape_evol}
\end{figure*}

The complexity of the GF98 and IF01 light curves and their mutual differences 
are better apparent in Fig.~\ref{pulseshape_evol}, which, for correspondence 
case ii), shows them  split into 9 panels, each displaying a single pulsation cycle,
with 125~ms time resolution.
The pulse shape of the IF01 pulsation and its evolution are apparent in the panels  
from 2 to 7. Two pulses and two dips nearly equally spaced are visible in the pulse
shape, with the second pulse stronger and fading later and almost suddenly in panel 7.
The pulse shape of the first 40~s of GF98 appears more complex than that of IF01. In the
same panels, it exhibits, from phase 0 to $\sim$~0.6, a single, broad pulse instead 
of the pulse and dip exhibited by IF01 in the same phase interval. The difference is
more marked in the first two and last two panels.

%
% Figure 5
%
\begin{figure}
\begin{center}
\centerline{\includegraphics[width=7cm]{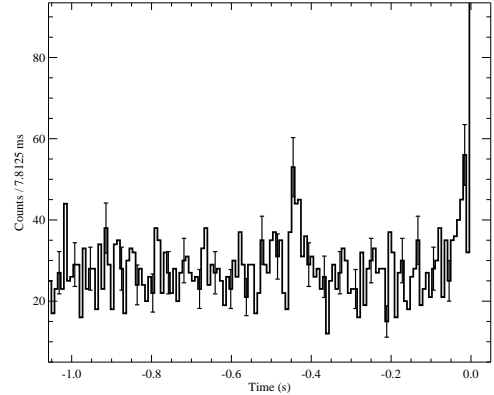}}
\caption{Precursor in the 40--700 keV light curve of GF98 at $t\sim -0.45$~s,
followed by a slow increase in intensity just before the initial pulse.
We added the counts of all GRBM units to improve the statistical
quality.}
\label{gf98_precursor}
\end{center}
\end{figure}
%

% Figure 6
%

%------------------------------
\subsection{Precursors}
%------------------------------
\label{s:prec}

The precursor of GF98, at $\sim 0.4$~s before the main event, is well
established. It was reported by 
Hurley et~al. (1999a)\nocite{Hurley99a} in the 25--150 keV energy band and by 
Mazets et al (1999b)\nocite{Mazets99b} in the 15--50 keV band.
In the 7.8125-ms GRBM data, we also find this precursor in the range 40--700 keV. 
It also occurs at $t = -0.45$~s and lasts about 0.1~s (see 
Fig.~\ref{gf98_precursor}). The pulse is even better apparent 
by rebinning the original 7.8125-ms data from all the 4 GRBM units in new 62.5-ms bins,
from which an excess of ($100\pm 23$~cts) can be established. From the same figure, 
it is also visible the slow increase in the count rate just before the initial spike, 
similar to the rise of a typical short burst (see also Mazets et al., 1999a).\nocite{Mazets99a}

\begin{figure}
\begin{center}
\centerline{\includegraphics[width=7.5cm]{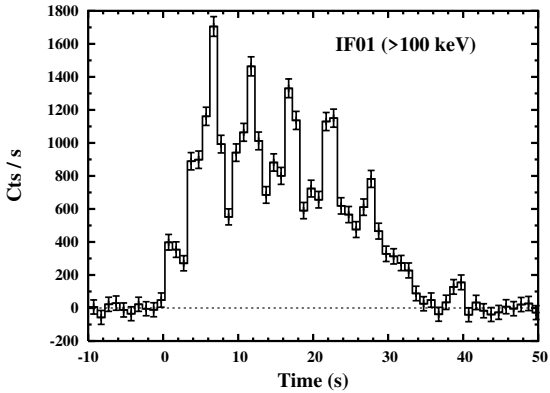}}
\caption{IF01 light curve in the $>100$~keV energy band. The fact that
the pulse preceding the main rise is also apparent in this band,
seems to suggest that it is unlikely to be a precursor.}
\label{f:precursor01}
\end{center}
\end{figure}

In the case of IF01, the nature of the first peak is more difficult to understand.
It may be a precursor candidate.
However its duration (about 3~s) is much longer than that
exhibited by the GF98 precursor, is not separated from the main rise of the flare,
and, more important, it shows a much harder spectrum than that of the
GF98 precursor. Indeed the latter was not detected by Konus \cite{Mazets99b} in
the energy channels above 50 keV, while the IF01 pulse is detected by the GRBM even
above 100 keV (see Fig.~\ref{f:precursor01}).

% Figure 7  
%

\subsection{Erratic time variability}

A temporal analysis of the 40--700 keV data with Fourier techniques was performed for 
each of the two flares. In the case of the 1998 flare, we limited the analysis to the 
first 38~s when the 5.2-s periodicity was not well set up yet.
We exploited both available time binnings of 7.8125~ms and, for the first 10~s, 
$\sim 0.5$~ms.

\begin{figure}
\centerline{\includegraphics[height=7.0cm,width=5.0cm,angle=270]{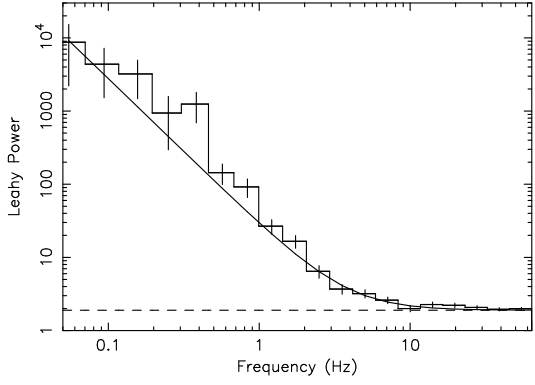}}
\caption[]{Measured PSD of IF01 using as time binning 7.8125~ms. {\it Solid line}: PSD
of the best-fit exponential function used to describe the mean light curve of IF01,
excluding the precursor (see text). {\it Dashed line}: Poissonian level corrected
for the dead time.}
\label{gf01_PSD}
\end{figure}

Using the longer time binning, for IF01  we estimated the PSD
function of the entire light curve in the 0.031--64~Hz frequency range. The resulting 
PSD, inclusive of the Poissonian variance, is shown in Fig.~\ref{gf01_PSD}.
We adopted the Leahy et al. (1983) normalization\nocite{Leahy83}, such that the 
Poissonian noise level has a PSD value of 2. As can be seen from this figure, apart from
two apparent peaks at 0.2 and 0.4 Hz due to the $\sim 5.2$-s periodicity, the main PSD
feature is its strong decrease with frequency. It mainly depends on the fact that 
the light curve we are analyzing is a non-stationary process.

%
% Figure 8
%

Indeed, in the null hypothesis that the light curve is a non-stationary Poisson process 
$x(t)$ of duration $T$ and mean value given by the deterministic function $\lambda(t)$,
the mean PSD of the process at frequency $f$, $\overline{S}_x(f)$, is
given by \cite{Frontera79}:
\begin{equation}
\overline{S}_x(f) = S_\lambda(f) + 2
\label{eq:PSD}
\end{equation}
where $S_\lambda(f)$ is the power spectrum of $\lambda(t)$ 
(Leahy et al. 1983 normalization adopted). 
Actually, the PSD of a non-stationary process can also be influenced by other effects, 
like the detector dead time ($\tau=4 \mu$s in our case), and the binning time
(coincident with the sampling time).  For low frequencies 
($f\ll f_\tau$ where $f_\tau = 1/(2\pi\tau)$), as in our case 
(Nyquist frequency $f_N= 64$~Hz), the dead time 
changes the  PSD level by a factor  $(1 - \mu\,\tau)^2$, where $\mu$ is the
average measured count rate (Frontera \& Fuligni 1978; see also van der Klis 1989).
\nocite{Frontera78,Klis89} 
The dead time effect, consistently with an expected correcting factor of 0.949 
(from the mean count rate $\mu=6412$~cts/s) is visible in the Poissonian level 
shown in Fig.~\ref{gf01_PSD}.

\begin{figure}
\centerline{\includegraphics[height=7.0cm,width=5.0cm,angle=270]{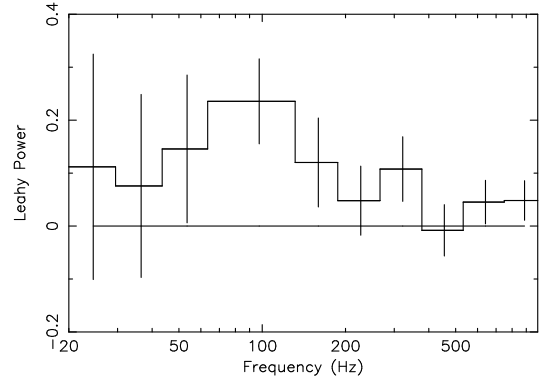}}
\caption[]{20~Hz--1~kHz PSD of IF01 measured in the time interval [+0.7,+10.7]~s after
subtraction of the non-stationary Poissonian noise dead-time corrected.}
\label{f:gf01_psd_sub}
\end{figure}

Following the working scheme adopted by Frontera \& Fuligni (1979), we have tested
the null hypothesis that the measured PSD can be entirely explained in terms of
a non-stationary process.
Assuming that the average behaviour of the IF01 light curve can
be described by an exponential $\lambda(t) = A U(t) e^{-k t}$, 
where $U(t)$ is the step function (time origin at the main rise 
of the flare, see discussion above), and $A$ and $k$ are free parameters,
the expected PSD is $S_\lambda(f) = 2 A^2/[\mu T (k^2 + (2\pi f)^2)]$.
From the fit to the light curve, we derived the best-fit parameters of the
exponential and thus the PSD of the function $S_\lambda(f)$.
After the addition of the Poissonian statistics corrected for the dead time, 
the expected PSD is shown in Fig.~\ref{gf01_PSD} as continuous line. 
As can be seen from this figure, $S_\lambda(f)$ dominates the PSD
up to $\sim 10$~Hz where it achieves the Poissonian level. We have found that
the power in excess of $S_\lambda(f)$ is negligible even above 10~Hz, with no  
evidence of a non-Poissonian noise up to 64~Hz. 
A similar analysis, performed in the first 10 s after the trigger using the 
high-time resolution binning of $\sim 0.5$~ms, has also given
a negative result up to 1~kHz (see Fig.~\ref{f:gf01_psd_sub}), with the conclusion 
that up to a frequency of 1~kHz no significant non-Poissonian noise is 
present in the time variability of IF01. 
 
%
% Figure 9
%

%
% Figure 10
%
\begin{figure}
\centerline{\includegraphics[height=7.0cm,width=5.0cm,angle=270]{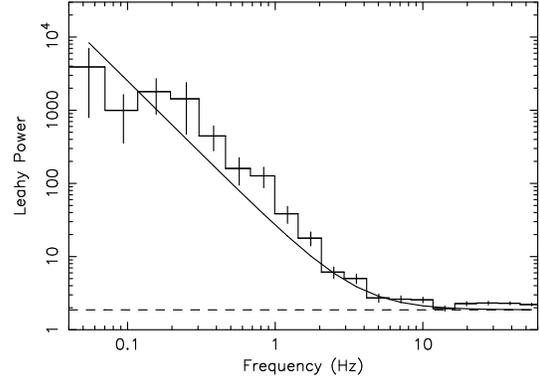}}
\caption[]{PSD of the first 38~s of GF98 with time origin 1~s after the onset,
in order to exclude the spike. The time binning used is 7.8125~ms. 
{\it Solid line}: PSD of the best-fit exponential function used to describe
the mean light curve. {\it Dashed line}: Poissonian level corrected
for the dead time.}
\label{f:gf98_38s_PSD}
\end{figure}

Following the same guidelines adopted above for IF01, we performed the PSD estimate 
of the first 38~s of GF98 after the initial spike.
The resulting PSD is shown in Fig.~\ref{f:gf98_38s_PSD} along with the $S_\lambda(f)$
(continuous line), obtained from the best-fit of the exponential function 
$\lambda(t)$ to the data, plus the Poissonian level corrected for the dead time.
As it can be seen, also in this case $S_\lambda(f)$ dominates the measured power 
spectrum up to $\sim$ 10~Hz, even if some evidence of an excess power over 
the non stationary Poissonian noise  is visible.

\begin{figure}
\centerline{\includegraphics[height=7.0cm,width=5.0cm,angle=270]{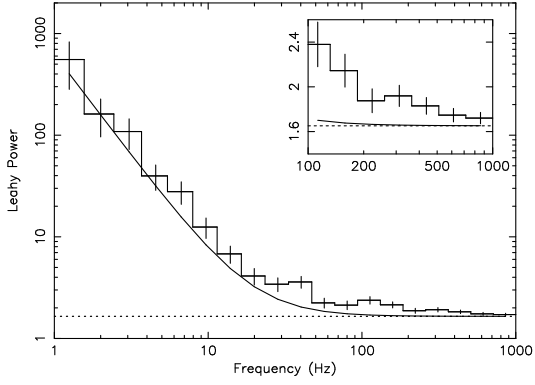}}
\caption[]{PSD of GF98 in the time interval [+1,+9]~s. {\it Continuum line}: 
PSD of the best-fit exponential function used to describe the 
first 10~s of the mean light curve. {\it Dashed line}: Poissonian level corrected
for dead time. {\it Inset}: zoomed high-frequency PSD.}
\label{gf98_pow_1-9_mod}
\end{figure}

To avoid a contamination of the PSD from the non-stationary component, we focused on 
the high-frequency domain deriving the PSD of the first 8~s after the spike,  
with high-time resolution binning  ($\sim$0.5~ms). The resulting PSD is shown in 
Fig.~\ref{gf98_pow_1-9_mod}, where it can be seen that, above 20~Hz, 
$S_\lambda(f)$ gives a small contribution to the total power and, above 100~Hz,
does not contribute at all (see also the inset in Fig.~\ref{gf98_pow_1-9_mod}).

%
% Figure 11
%

\begin{figure}
\centerline{\includegraphics[height=7.0cm,width=5.0cm,angle=270]{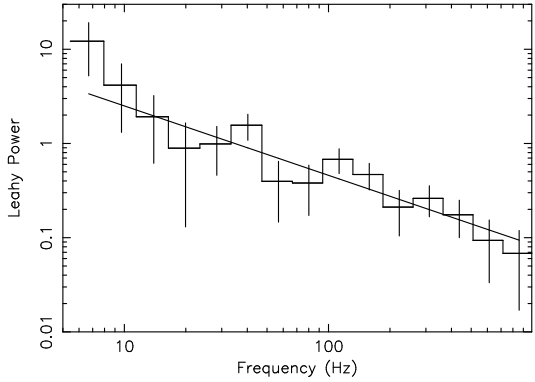}}
\caption[]{PSD of the residual noise of GF98 in the time interval [1, 9]~s.
{\it Solid line}: best-fit power-law (see text).}
\label{gf98_pow_1-9_sub_fit}
\end{figure}

\begin{figure}
\centerline{\includegraphics[height=7.0cm,width=5.0cm,angle=270]{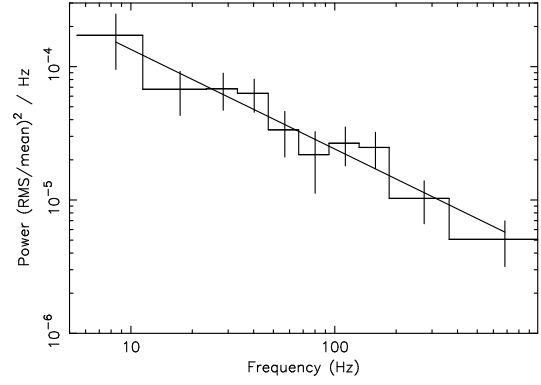}}
\caption[]{PSD of GF98 for the interval [+1,+9]~s averaged over the
four GRBM units. {\it Solid line}: best-fit power law (see text).}
\label{gf98_fracrms_ave_grp2_fit}
\end{figure}

The excess power with respect to the non-stationary Poisson model in the 6--1000~Hz
for the GRBM unit 1 is shown in 
Fig.~\ref{gf98_pow_1-9_sub_fit}. As can be seen, this time a relevant residual power 
is apparent up to 1~kHz. The non--Poissonian spectrum $S_{np}(f)$ in the range 10-1000 Hz 
is well fit with a power law ($S_{np}(f) \propto f^{-\alpha}$) 
with $\alpha = 0.74 \pm 0.18$. 
An excess power has also been found in the PSD of the other GRBM units
(2, 3, and 4). The cumulative result is shown in Fig.~\ref{gf98_fracrms_ave_grp2_fit}.
A power-law still gives the best description of the non--Poissonian spectrum with a
best-fit power-law index of $0.75\pm 0.15$, even though also a flicker noise 
($S_{np} \propto 1/f$) gives an acceptable fit ($\chi^2/{\rm dof} = 10.6/9$).
Using the power-law best-fit values, the total fractional variation (in RMS units) 
in the range 10--1000~Hz comes out to be around 1\%.
Concerning IF01, as far as we assume similar fractional variability, we could not
detect it because of the worse statistics; therefore, we cannot rule out the
presence of similar non--Poissonian noise with comparable power for IF01.

%
% Figure 12
%

\subsection{Spectral properties}
\label{s:spectra}

\begin{figure}[t]
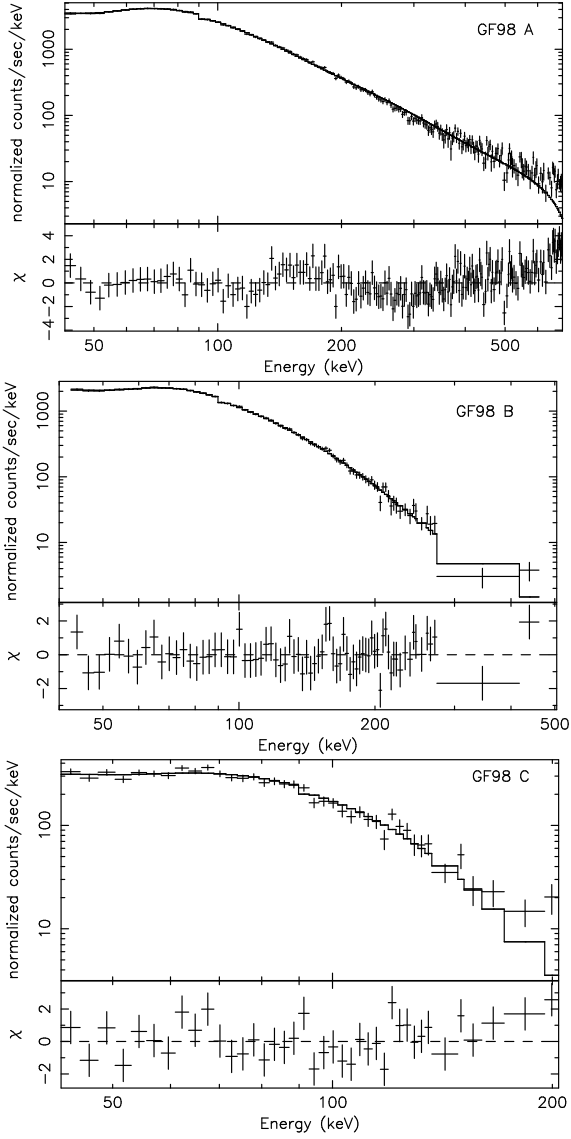

\begin{center}
\centerline{\includegraphics[width=5.0cm,angle=270]{0511_15a.ps}}
\centerline{\includegraphics[width=5.0cm,angle=270]{0511_15b.ps}}
\centerline{\includegraphics[width=5.0cm,angle=270]{0511_15c.ps}}
\caption{Count rate spectra of GF98 and their fit with a {\sc bb + bknpl} during
the interval A and B, and with a {\sc bb} model during the interval C.
The residuals to the model are shown as well.}
\label{f:spectra_98}
\end{center}
\end{figure}

The two available  source count rate spectra  (A and B) of IF01 and the three (A, B, 
and C)
source spectra of GF98 are shown in Fig.~\ref{f:spectra_01} and Fig.~\ref{f:spectra_98},
respectively. For GF98, the A spectrum (inclusive of the initial spike) is well 
determined up to 700 keV, the B spectrum up to $\sim 500$ keV and the C spectrum up to 
200 keV. 
As far as IF01 is concerned, the  spectrum A is well determined up to 700 keV, while 
the B spectrum can only be estimated up to 100 keV. In the following, we limited our 
analysis to the energy bands where significant source counts were detected.

%
% Figure 13
%

%
% Figure 14
%
\begin{figure}[t]
\begin{center}
\centerline{\includegraphics[width=5.0cm,angle=270]{0511_14a.ps}}
\centerline{\includegraphics[width=5.0cm,angle=270]{0511_14b.ps}}
\caption{Count rate spectra of IF01 and their fit with a {\sc bb + bknpl} during
the interval A, and with a {\sc bb} model during the interval B. 
The residuals to the models are shown as well.}
\label{f:spectra_01}
\end{center}
\end{figure}
%

%
% Figure 15
%

\begin{figure}[t]
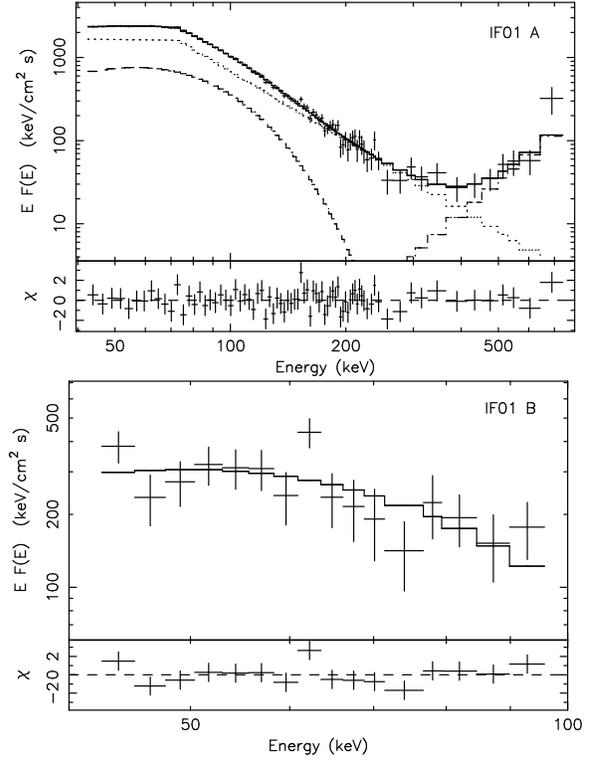

\begin{center}
\centerline{\includegraphics[width=5.0cm,angle=270]{0511_16a.ps}}
\centerline{\includegraphics[width=5.0cm,angle=270]{0511_16b.ps}}
\caption{$E F(E)$ spectra of IF01. {\em Top panel}: A spectrum along with the
  best fit with a {\sc bb+bknpl+pl} model. {\em Bottom panel}: B spectrum  along with
  the fit with a {\sc bb}. Also the single model components and the
residuals (in units of $\sigma$) to the best-fit models are shown.}
\label{gf01_spectra}
\end{center}
\end{figure}

Several single, two--component and three--component  models were tested 
to fit these spectra. In the case of the A spectrum of IF01,
models like a power-law {\sc pl}, a cutoff power law ({\sc cutoffpl}), a single {\sc bb},
an {\sc ottb} with or without a  power-law {\sc pl}, are unsatisfactory,
even in the 40--300 keV energy band. 
Up to 300 keV, a {\sc bknpl} plus either a {\sc bb} or an {\sc ottb}  give a good fit
($\chi^2/{\rm dof} =56/65$ with {\sc bb} and $\chi^2/{\rm dof} =61/65$ with {\sc ottb}).
Also the sum of two {\sc bb} cannot be ruled out ($\chi^2/{\rm dof}=72.6/66$).
In the top panel of Fig.~\ref{f:spectra_01} we show the fit of the A spectrum
with the {\sc bb} plus {\sc bknpl} model.
Clearly, an excess to the model is apparent in the 300--700 keV band.
The best fit of the entire spectrum is obtained when a {\sc pl} is added to either
a {\sc bb + bknpl} model or an {\sc ottb + bknpl} model.
The best-fit parameters along with the $\chi^2$ values are reported in 
Table~\ref{tab:spectra}, while in Fig.~\ref{gf01_spectra} (top) the $EF(E)$ spectrum 
along with one of the best models ({\sc bb + bknpl + pl}) and the 
residuals to the model are shown.
By adding a {\sc pl} to the double {\sc bb}, the fit goodness 
($\chi^2/{\rm dof}=83/74$) is also acceptable, even though worse than the fit with the 
two previous models, with the following best-fit parameters: 
$kT_{bb1}=13.5^{+0.2}_{-0.2}$~keV, $kT_{bb2}=33^{+2}_{-3}$~keV, and
$\Gamma=-1.06^{+0.64}_{-0.03}$.

%
% Figure 16
%

The B spectrum (derived up to only 100 keV)  can be fit with either a {\sc bb} or 
an {\sc ottb} ($\chi^2/{\rm dof} = 17.6/14$ and $\chi^2/{\rm dof} = 16.7/14$, 
respectively). The best-fit parameters are also reported in Table~\ref{tab:spectra}, 
while the $E F(E)$ spectrum and its best-fit model are shown in Fig.~\ref{gf01_spectra}
(bottom panel). The multi-component models adopted for the A spectrum also yield
acceptable fits of this spectrum.

% Table 2
%
\begin{table*}
\begin{center}
\caption{Best-fit parameters of the IF01 and GF98 photon spectra.}
\begin{tabular}{ccccccccccc}
\hline
\noalign{\smallskip}
Flare &  Interval & Band & Model & $kT$ &  $\Gamma_1^{\rm (a)}$ &  
   $\Gamma_2^{\rm (a)}$ & $E_{\rm break}^{\rm (a)}$ &  $\Gamma^{\rm (b)}$ & $\chi^2/{\rm dof}$
   & Fluence$^{\rm (c)}$  \\
      &        &  (keV)              &     & (keV)         &             &       
      & (keV)  &           &   &  ($10^{-4}$~erg cm$^{-2}$) \\
\noalign{\smallskip}
\hline
\noalign{\smallskip}
IF01  & A      & 40--700 & {\sc bb+bknpl+pl} & $14.4^{+1.2}_{-0.8}$ & $2.1^{+0.1}_{-0.1}$ &
    $4.8^{+0.2}_{-0.2}$ & $73^{+2}_{-2}$ & $<-0.4$ & 62/72 
      & 1.2\\
      &        &          & {\sc ottb+bknpl+pl} & $31.7^{+2.3}_{-2.1}$ & $0.6^{+0.2}_{-0.1}$ & 
    $5.0^{+0.3}_{-0.2}$ & $73^{+1}_{-2}$ & $-0.6^{+1.5}_{-0.6}$ & 64/72
      & 1.2 \\
      & B      & 40--100  & {\sc bb} & $12.7^{+1.6}_{-1.4}$ & -- &  -- & -- & -- & 17.6/14
      & 0.03\\
      &        &           & {\sc ottb} & $36^{+12}_{-8}$ & -- &  -- & -- & -- & 16.7/14
      & 0.03 \\
\noalign{\smallskip}
 GF98 & A  & 40--700 & {\sc bb+bknpl+pl} & $19^{+4}_{-3}$ & $1.0^{+0.2}_{-0.3}$ & 
     $3.7^{+0.1}_{-0.1}$ & $75^{+1}_{-2}$ & $<-0.5$ & 191/187 
      & $> 6.4$ \\
      &    &         & {\sc ottb+bknpl+pl} & $40^{+3}_{-3}$ & $-1.8^{+0.2}_{-0.6}$ & 
     $3.48^{+0.02}_{-0.02}$ & $71^{+2}_{-1}$ & $-2.1^{+1.1}_{-0.4}$ & 191/187 
      & $> 6.4$\\
      & B & 40--500 & {\sc bb+bknpl} &  $20.5^{+1.8}_{-1.8}$ & $1.3^{+0.2}_{-0.3}$ & 
     $4.9^{+0.2}_{-0.1}$ & $75^{+2}_{-2}$ & -- & 50.3/69 & 3.1 \\
      &   &         & {\sc ottb+bknpl} &  $44^{+3}_{-2}$ & $<-1.0$ & 
     $4.9^{+0.6}_{-0.3}$ & $72^{+1}_{-1}$ & -- & 50.7/69 & 3.1 \\
      & C & 40--200 & {\sc bb+pl} & $16.0^{+0.4}_{-0.4}$ & -- &
      -- & -- & $<2.6$  & 41.3/36 & 0.49\\
\noalign{\smallskip}
\hline
\noalign{\smallskip}
\multicolumn{11}{l}{$^{\rm (a)}$ Parameters of the broken power law: $I(E)\propto
E^{-\Gamma_1}$ for $E\le E_{break}$;
$I(E)\propto E^{-\Gamma_2}$ for $E\ge E_{break}$} \\
\multicolumn{11}{l}{$^{\rm (b)}$ Parameters of the power law: $I(E)\propto E^{-\Gamma}$} \\
\multicolumn{11}{l}{$^{\rm (c)}$ Fluence in the 40--700 keV energy band.}

\end{tabular}
\label{tab:spectra}
\end{center}
\end{table*}

As far as the GF98 spectra are concerned, first of all we caution that
the results of the fits of the A spectrum can be can be influenced by the 
dramatic dead-time and pile-up effects due to the hard initial spike contribution, 
the hardness of which rapidly changes with time (see Mazets et~al. 1999b and also F99).

\begin{figure}[t]
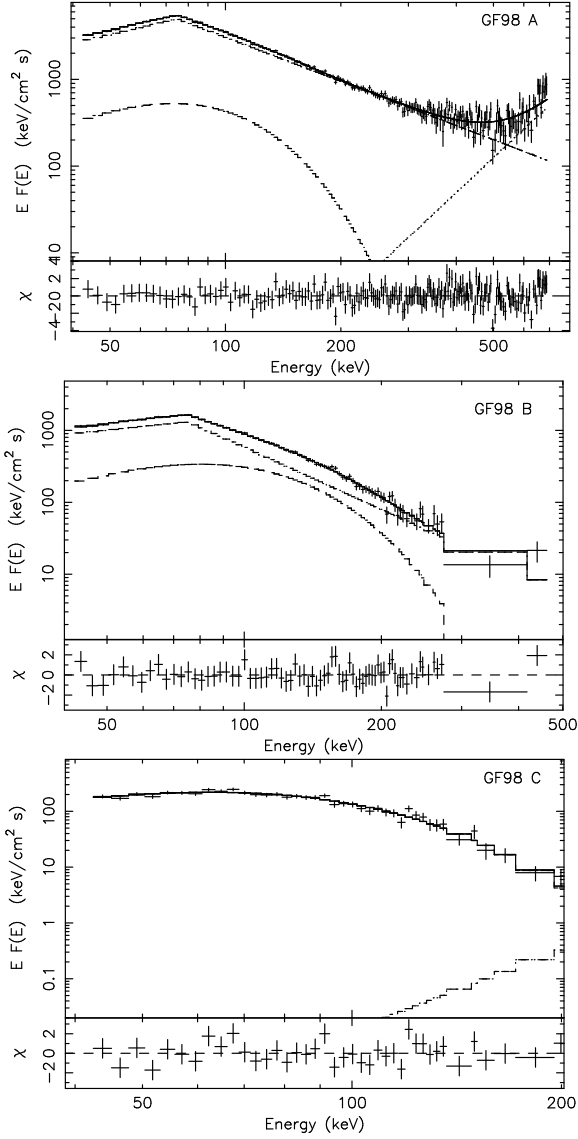

\begin{center}
\centerline{\includegraphics[width=5.0cm,angle=270]{0511_17a.ps}}
\centerline{\includegraphics[width=5.0cm,angle=270]{0511_17b.ps}}
\centerline{\includegraphics[width=5.0cm,angle=270]{0511_17c.ps}}
\caption{$E F(E)$ average spectra of GF98. {\it Top panel}: A spectrum with superposed 
the best-fit model {\sc bb+bknpl+pl}. {Middle panel}: B spectrum with superposed the
best-fit model {\sc bb+bknpl}. {\it Bottom panel}: C spectrum with superposed
the best-fit model {\sc bb+pl}. Also the model components and the
residuals to the best-fit models are shown.}
\label{gf98_spectra}
\end{center}
\end{figure}

In spite of that, the best fit of the A spectrum is obtained using as input the 
three--component models which best fit the A spectrum of IF01:
either a {\sc bb} plus {\sc bknpl} plus {\sc pl} or an {\sc ottb} plus 
{\sc bknpl} plus {\sc pl}. 
The best-fit parameters along with the $\chi^2$ values are also reported in 
Table~\ref{tab:spectra}, while
in Fig.~\ref{gf98_spectra} (top) the $EF(E)$ spectrum along with one of the
best models ({\sc bb + bknpl + pl}) and the residuals to the model are shown.
The simple {\sc bb + pl} or the {\sc ottb + pl} models,
which were found to provide a good fit of the same spectrum with the preliminary
GRBM response function and the 10\% systematic error adopted by F99, with the improved
response function provide unsatisfactory fits ($\chi^2/{\rm dof} > 2$ with
${\rm dof} = 189$).

% 
% Figure 17
%

The fit of the B spectrum was limited to the 40--500 keV energy band.
The best-fit model is obtained  with either a {\sc bb+bknpl} 
($\chi^2/{\rm dof} = 50.3/69$)  or an {\sc ottb+bknpl} ($\chi^2/{\rm dof} = 50.7/69$).
However, in the latter case, the lower limit to the photon index $\Gamma_1$ cannot be 
constrained (see Table~\ref{tab:spectra}).
In agreement with the results by F01, we find that the {\sc ottb + pl} model,
the  {\sc bb + pl} and the Band model \cite{Band93}  are definitely unacceptable 
(e.g., $\chi^2/{\rm dof}= 860/70$ for the former model).
However, we find that also the double {\sc bb} plus {\sc pl} model, which was found by 
F01 to provide the best fit to the GRBM plus {\it Ulysses} spectral 
data, does not provide a good fit to the 40--500 keV GRBM data alone with the improved
response function ($\chi^2/dof=99/69$). 
The best-fit parameters are reported in Table~\ref{tab:spectra}, while 
in Fig.~\ref{gf98_spectra} (middle panel) we show the $E F(E)$ spectrum with superposed
the best-fit {\sc bb + bknpl} model  and the residuals to the model.

The C spectrum could be estimated in an even more restricted energy band (40--200~keV),
due to the poor statistics above 200~keV.
The fit with a simple {\sc bb} gives a $\chi^2/{\rm dof} = 52/38$, with a chance 
probability of 6.5\%. By adding a {\sc pl} component, the fit improves 
($\chi^2/{\rm dof} = 41.3/36$) even if the photon index cannot be strongly
constrained ($<2.6$ at 90\% CL). However, this component is needed to 
model the high-energy excess with respect to the {\sc bb} component.  
The best-fit parameters of the {\sc bb + pl} model are reported in 
Table~\ref{tab:spectra},
while the $E F(E)$ spectrum along with the best-fit model and the residuals to the
model are shown in Fig.~\ref{gf98_spectra} (bottom panel).

The 40--700 keV fluence in each of the time intervals is shown in Table~\ref{tab:spectra}. 
For the interval A of GF98 we have estimated only the lower limit of
the 40--700 keV fluence due to the high dead-time and pile-up 
effects during the initial spike.

The contribution to the fluence from the {\sc bknpl} component is the most relevant 
in most of the time of the GF98 event and for almost the entire time duration of the 
IF01 (see Figs.~\ref{gf01_spectra} and \ref{gf98_spectra}).
In the case of GF98, assuming {\sc bb} as thermal component, the fractional 
contribution of the {\sc bknpl} component is 86\%  in the interval A and 75\% in 
the interval B, while it is 68\% in the time interval A of IF01. 
These values decrease if the {\sc bb} is replaced by an {\sc ottb}.
Instead the fractional contribution of the high-energy {\sc pl} component, which is 
apparent only in the time interval A of both GF98 and IF01 (and, perhaps, in GF98 C
as well), is only 2\%. It is 
interesting the fact that this component, which in the case of GF98 could be attributed 
to the initial spike, is also present even in the case of IF01, which does 
not exhibit any spike.

The total 40--700 keV fluence of the IF01 event is $S_{01} = 
1.2 \times 10^{-4}$~erg~cm$^{-2}$ to be compared with a value $S_{98} = 1 
\times 10^{-3}$ erg~cm$^{-2}$ derived for GF98, which must be considered a lower 
limit of the real fluence as explained above.

%-------------------------------------------------
\subsection{Spectral Evolution}
%-------------------------------------------------

Given that only two energy channels are available on 1-s integration time, the
only way to investigate how the spectrum evolves with time is the ratio
between the counting rates in the 100--700~keV channel and those in the 40--100~keV
channel (Hardness Ratio, HR). However, since the two flares occurred at different
directions with respect to the BeppoSAX local frame, the differences in the HR
might be at least partially ascribed to the different instrumental response.
Because of this, in order to compare the absolute hardnesses,
we studied the equivalent $kT$ of an {\sc ottb} model, although in principle it does
not provide us with an acceptable fit of the time-averaged spectra.

\begin{figure}[t]
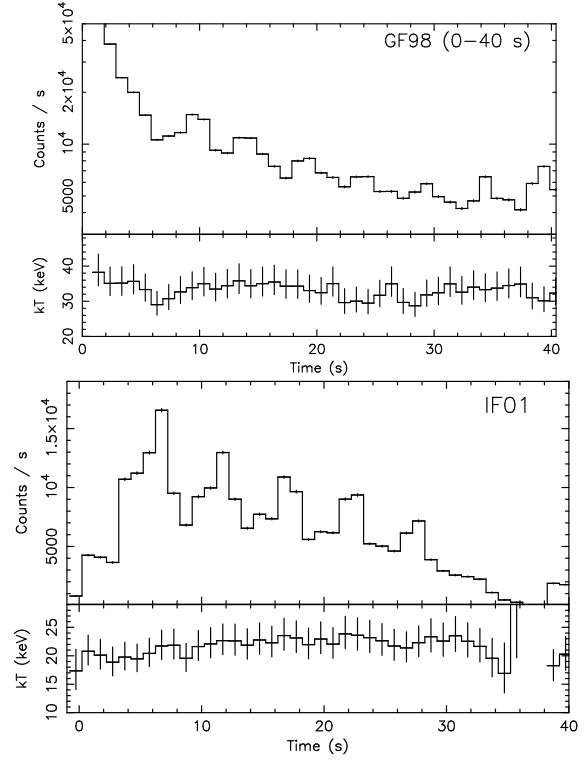

\begin{center}
\centerline{\includegraphics[width=5.0cm,angle=270]{0511_18a.ps}}
\centerline{\includegraphics[width=5.0cm,angle=270]{0511_18b.ps}}
\caption{Time behaviour of the equivalent $kT$ of an {\sc ottb} model
for the entire duration of IF01 and for the early part of GF98. Error bars
are 90\% CL. {\it Top panel}: GF98; {\it bottom panel}: IF01.}
\label{f:gf98_if01_kT}
\end{center}
\end{figure}

The time behaviour of $kT$ for the entire duration of GF98 is already reported (F01).
Limiting the $kT$ of GF98 to the first 40~s, for a comparison with IF01 
(see Fig.~\ref{f:gf98_if01_kT}), we see that the $kT$ of IF01 is significantly lower 
than that of GF98. In addition it does not exhibit any significant variations throughout 
the event. In particular, there is no clear indication of correlation between $kT$ and the 
modulation observed in the flare profile. However, a slight trend of the $kT$ to increase
with time and then to decrease before the end of the event is apparent.

% 
% Figure 18
%
%
%

%%%%%%%%%%%%%%%%%%%%%%%%%%%%------------------------------
\section{Discussion}
%%%%%%%%%%%%%%%%%%%%%%%%%%%%------------------------------
The two large flares were classified on the basis of their energetics and durations
as giant (GF98) and intermediate (IF01) \cite{Kouveliotou01}.
We find this classification appropriate when their 40--700~keV fluence is considered
($S_{98} > 1 \times 10^{-3}$ erg~cm$^{-2}$, $S_{01} = 1.2 
\times 10^{-4}$ erg~cm$^{-2}$), which give a ratio of $>10$.
However, in the next sections we discuss other similarities and distinctive
features of the two events on the basis of their spectra and temporal variability.

\subsection{Light curves}
For the duration of IF01 ($\sim 40$~s), both light curves of GF98 and IF01 
are characterized for most part of the time by a $\sim 0.2$~Hz pulsation, with complex
and variable shape of the periodic pulses. The pulse shapes are different in the two
cases. Only dips between pulses appear the common feature
of both light curves. The most distinguishing feature of the GF98 light curve
is the presence of an initial spike, which is not observed in the case of IF01. 
In fact, given the difficulty of justifying the first weak pulse in the IF01
light curve as a precursor (see Section~\ref{s:prec}), it appears that the initial
GF98 spike corresponds to this pulse (if the two events are assumed to be
``phased'' by their dips). This is one of the most important results
of our comparative analysis. 
    
The periodicity found in the IF01 light curve ($P_{01} = 5.21 \pm 0.05$~s)
is consistent with that measured during the source quiescence before and soon after the 
event ($P_{quiesc} = 5.17284268$~s) \cite{Woods03b}.

Within the magnetar model scenario (see F01, Thompson \& Duncan 2001),
\nocite{Thompson01} the 1998 giant flare
is triggered by a distortion of the internal magnetic field in the neutron star
core, that induces large-scale fractures in the crust and strong magnetic
shears in the magnetosphere, that drive reconnection and conversion to Alf\'ven waves.
According to this view, the initial spike is the signature of a relativistic outflow 
with a very low baryon load, as also suggested by the radio transient observed 
by Frail et~al. (1999)\nocite{Frail99} and corroborated by the highly structured temporal 
profile of the spike, with peaks as narrow as $\sim10^{-2}$~s \cite{Mazets99b},
whereas the pulsating tail would be due to the fraction (about 50\%) of energy trapped
in the magnetosphere in the form of a photon-pair plasma. 
A similar scenario for the IF01 time profile is clearly problematic: we
do not see the spike. The spike absence could be explained in several ways.
We focus on three possibilities: first, no huge energy release comparable to that of 
GF98 occurred; second, the spiking event occurred, but the beamed outflow was 
not directed toward our line of sight; third, a comparable or slightly less energetic 
release really occurred, but the permanent changes undergone by the magnetosphere 
after the global reconfiguration further to GF98 \cite{Woods01} are responsible for 
the unusual time profile of IF01.
The first explanation seems the most natural, since it easily accounts for the
lower X-- and $\gamma$-ray fluence of IF01 and seems to agree
with the minor changes observed in the trend of the pulsation profile after the flare
against what occurred after GF98
\cite{Gogus02,Woods03b}. However the light curve of  IF01 shows some similarities
(e.g., dips at the same phase in the correspondence case of
Fig.~\ref{both_offsets_1900_top_interpulse}) to the initial stage of GF98.
Furthermore, the pulse profile during IF01 appears far more complex
than it appeared at the later stages of GF98 \cite{Mazets99b} and than that
of the quiescent pulsar after 1998 \cite{Gogus02,Woods03b}.
This suggests that a transient reconfiguration of the magnetic
field took place related to this event, that caused the complex pulse
profile evolution during the burst.
In addition, the detection of a non-thermal X-ray afterglow after this event
raises the issue of a possible GRB-like mechanism for explaining
this emission \cite{Feroci03,Ioka01}, that in turn would imply an
outflow of relativistic particles (whose signature is missing in IF01, however).

\begin{figure}[t]
\centerline{\includegraphics[height=7.5cm,angle=270]{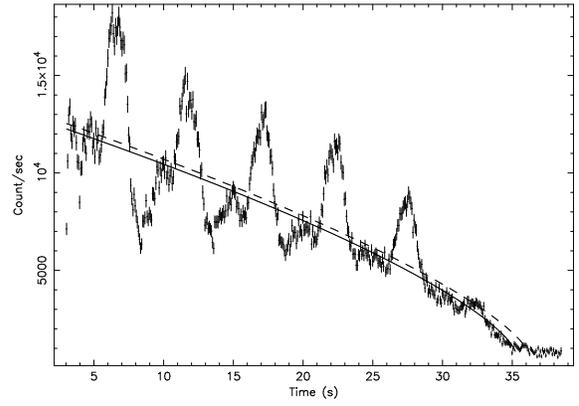}}
\caption[]{Light curve of IF01; the dashed line is the best
fit with a fireball model obtained for the time interval [7.46875, 38.5]~s,
rebinned at $5.171875$~s; the solid line shows the best fit applied
to the interval [3, 38]~s, whose parameters have been constrained
to vary within 1$\sigma$ around the best fit of the tail.}
\label{f:fireball}
\end{figure}

In the context of the magnetar model, F01 pointed out that the 
envelope of the light curve can give an important clue  about
the radiation emission mechanisms and/or the geometry during the flares.
Thompson and Duncan (2001),\nocite{Thompson01}
assuming that the emitted luminosity is the result of a cooling fireball
trapped on the closed magnetic field lines of a neutron star, expect that it
varies as a power of the remaining fireball energy $E^a$. As a consequence the
fading law of the radiation is expected to vary as 
$L(t) = L(0) (1-t/t_{evap})^{a/(1-a)}$, where $t_{evap}$ is the time at which
the fireball evaporates and its radius shrinks to zero, while $a$
depends on the trapped fireball geometry and temperature distribution ($a=2/3$
or $a=1/2$ in the case of spherical or cylindrical geometry, respectively).
F01 found, for GF98, $t_{evap}=(501\pm 13)$~s and $a= 0.828\pm0.005$ in
the 40--100 keV range, and $t_{evap}=(545\pm 62)$~s and $a= 0.85\pm0.01$
in the 100--700 keV band.
We find for the 40--700 keV the result shown in  Fig.~\ref{f:fireball} with
the following best fit parameters: $t_{evap}=(35.25\pm0.11)$~s, $a= 0.413\pm0.004$.
This result shows that, in contrast to the spherical-like geometry of GF98,
the trapped fireball responsible for the pulsating tail of IF01 had probably
a cylinder-like geometry ($a\sim 1/2$), with a non uniform temperature
distribution (that has the effect of decreasing the value of the fireball
index $a$).

% Figure 19
%++++++++++++++++++
%++++++++++++++++++

\subsection{Erratic time variability}
From the PSD estimates, it clearly emerges that GF98 has time properties which are
not observed in the case of IF01.
A significant power above the Poissonian level is apparent up to 1~kHz 
in the PSD of GF98, with a power-law shape (index $\alpha = 0.75 \pm 0.15$) with 
frequency.  Even if the 10--1000 Hz fractional variation (in RMS units) 
is only 1\%, the PSD shows the clear presence of a flicker--like noise, similar to
that observed in many accreting compact X--ray sources. 

Barat et al. (1983)\nocite{Barat83} found evidence of timing noise in the X-ray
decay of the March 5 burst from  SGR0526-66, although in that case it was
identified as due to $\sim23$~ms quasi-periodic oscillations.
An interpretation of the flicker--like noise in the light of the magnetar model
is needed, although it could be possibly explained invoking the mechanisms
described by Duncan (1998) \nocite{Duncan98} and by Ioka (2001): \nocite{Ioka01}
they account for the ms quasi-periodicity as due to the excitation
of global seismic oscillations of the crust coupling strongly to Alf\'ven modes
in the lower magnetosphere.

A possible phenomenological interpretation of the detected noise is in terms of
clustering of elementary pulses. In this case the noise dependence on frequency is
determined by the distribution of waiting times between subsequent pulses
and, possibly, by the distribution of the pulse intensity.
There is a big variety of mechanisms that, in principle, may generate power-law
noise with $\alpha\approx1$: see, e.g., Kaulakys \& Me\u{s}kauskas (1998) \nocite{Kaulakys98}
and references therein.
Davidsen \& Schuster (2002)\nocite{Davidsen02} propose a simple mechanism for 
generating pulse sequences with $S(f) \propto f^{-\alpha}$ in systems whose dynamics 
is driven by a variable threshold, like for earthquakes. This kind of mechanism is 
also known as an integrate-and-fire (IAF) model. It requires a threshold $C(t)$ evolving 
with time according to a Brownian motion with diffusion constant $D$ within a
defined interval $C_l<C(t)<C_u$. Assuming a function $V(t)$ which linearly increases
with time, as soon as $V(t)$ matches the threshold $C(t)$, it is reset
to a starting value $V_0$ and a single pulse is produced, leaving $C(t)$
unaltered. Under proper choices of the threshold boundaries, diffusion constant
and reset value, the PSD of the resulting pulse train is characterized by
a PSD with a power-law shape with index $\alpha$, which is typically
between 0.5 and 1.1. The similarity of the high-frequency noise with that of
earthquakes was also discussed by Cheng et~al. (1996)\nocite{Cheng96}, who showed that
for a set of bursts from SGR1806-20 the distribution of size and
cumulative waiting times are similar to those of earthquakes. These properties
have also been verified for the SGR1900+14 short bursts \cite{Gogus99}.

In the light of the magnetar model, the observed noise could be the consequence of the
dramatic energy release during the initial spike. Likely, the engine which powered 
the spike is far from equilibrium and the magnetic field
lines probably undergo rapid and complex variations. If a threshold-controlled
mechanism, like that above described, is at work, unstable 
fireballs could be created, that burst whenever a particular threshold is exceeded,
that might depend on quantities like energy density, magnetic field and its twist. 
In this scenario, the observed high-frequency noise might be suggestive of how 
fractures in the crust propagate with time.

\subsection{Energy spectra}
At first glance, we notice that the GF98 spectrum of the
first 40~s is significantly harder than that of IF01, as shown by the comparison
of the equivalent temperatures of an {\sc ottb} model.
However, unlike the discussed differences between the two events, the spectral properties
of GF98 and IF01 show striking similarities. In the time interval A, which
covers most of the IF01 light curve and during the first 68~s of the GF98 light curve,
both 40--700 keV spectra are best fit with the same three-component models
({\sc bb + bknpl + pl} or {\sc ottb + bknpl + pl}; see Table~\ref{tab:spectra}).
Assuming the former of these models, both the {\sc bb} temperatures and break energies of 
the {\sc bknpl} model are similar: $kT_{bb}^{98} = 19^{+4}_{-3}$ keV and 
$E_{break}^{98} = 75^{+1}_{-2}$ keV, $kT_{bb}^{01} = 14.4^{+1.2}_{-0.8}$ keV and 
$E_{break}^{01} = 73^{+2}_{-2}$ keV, respectively.  However the centroids  
of the {\sc bknpl} indices are only marginally consistent at 90\% CL,
and we remind that the A spectrum of GF98 is likely  distorted
by dramatic dead-time and pile-up effects suffered during the initial spike.

The {\sc bb + bknpl (+pl)} model better agrees with the trapped fireball+corona scenario
\cite{Thompson01},
at least in the case of GF98: the {\sc bb} flux remains almost constant throughout the 
flare, whereas the {\sc bknpl} and {\sc pl} fade exhibiting a small spectral evolution.
This is confirmed also by the temporal evolution of the equivalent $kT$:
while GF98 exhibits a slow decrease during the
first 40~s, after which it softens even more slowly, IF01 shows a mild increase
followed at the end by a small drop, with no strong spectral evolution.
In this scenario, the {\sc bb} component is due to the outer layer of a trapped fireball,
while the {\sc bknpl} can partially come from the surrounding corona, and,
probably, from the reprocessing of the radiation coming from the inner
fireball. Indeed such component, although it decreases, does not disappear in the time
interval B of GF98, where the corona should already have evaporated.
The {\sc bb} temperature value in both GF98 and IF01 are above the minimum photospheric
temperature of a trapped fireball expected in \mbox{$B>B_{\textrm{QED}}$}
magnetic fields (eq.~133 from Thompson \& Duncan 1995).\nocite{Thompson95}

While a possible interpretation can be given for the {\sc bknpl}, the origin of
the high-energy ($>300$ keV) power-law component, with positive index at least
for the A spectra of both events, is more mysterious.

\subsection{The last isolated pulse of IF01}
The 1.5-s long isolated pulse of IF01 has a shape apparently different 
from that of the previous pulses. It is narrower and is not in phase with the previous
ones. Thus, we are led to think about a different origin for it. A possibility is
that, unlike the periodic light curve before it, whose likely origin is
in the outer layer of the transient corona, the last pulse could originate in the nearby 
of the neutron star surface and/or  with the same mechanism of typical short bursts. 
We note that, the occurrence of short bursts during the tail of bright bursts
was noted also in the tail of the August 29 event \cite{Palmer02}.
Actually, its average energy spectrum does not appear significantly different from
the A spectrum, and Fig.~\ref{f:gf98_if01_kT} shows that its $kT$
is consistent with general behaviour at previous times.
Thus, the last pulse mainly differs from the first part of the flare for its
temporal properties rather than for its energy spectrum.

% Fig 20
%
\begin{figure}
\begin{center}
\centerline{\includegraphics[width=8cm]{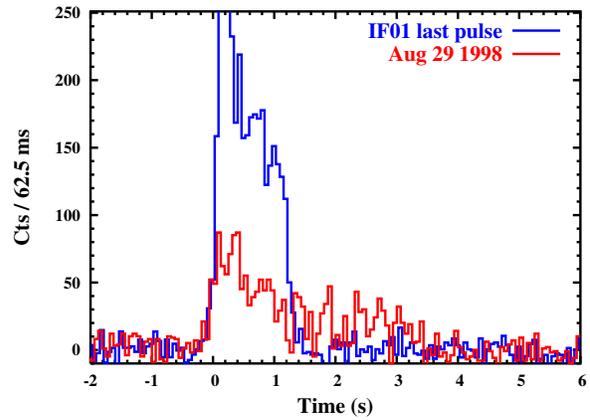}}
\caption{40--700~keV time profiles of the last isolated pulse of IF01 ({\em blue})
and of the August 29, 1998 burst ({\em red}).}
\label{gf01_lastpulse_980829}
\end{center}
\end{figure}
Actually, if we compare this pulse  with the peculiar burst of 
August 29, 1998 \cite{Ibrahim01}, also detected with the GRBM (see 
Fig.~\ref{gf01_lastpulse_980829}), we find a similar light curve, even if
the fluence of the IF01 last pulse ($\sim 5\times 10^{-6}$~erg cm$^{-2}$)
is about twice as high as that of the August 29 burst fluence in the 40--700~keV band.
Curiously, the duration of the subpulse of the August 29 burst is similar to that
of the last pulse of IF01.
This similarity could suggest a common origin of the last pulse of IF01
with some of the brightest short bursts recorded from SGR1900+14.

%%%%%%%%%%%%%%%%%%%%%%%%%%%%------------------------------
\section{Conclusions}
%%%%%%%%%%%%%%%%%%%%%%%%%%%%------------------------------
The \sax\ GRBM was the only instrument  that in the same 40--700 keV energy band 
allowed a detailed study of the two  large X--/gamma--ray flares from SGR1900+14
occurred on 1998 August 27 and 2001 April 18.
In this paper we have compared the spectral and temporal properties of both
flares to study their similarities and distinctive features. 
Apart from the different time durations of the two flares ($\sim 40$~s for
IF01 and $\sim 300$~s for GF98) and a higher ($>10$) 40--700 keV fluence 
of GF98,  other distinctive features have also
been derived from our analysis. The light curve of IF01 does not show the
initial spike exhibited by GF98, and shows a periodicity ($5.21 \pm 0.05$~s) 
consistent with the value measured for the quiescent X-ray source
soon after IF01 \cite{Woods03b}, with no evidence for a glitch like in
the case of GF98 \cite{Woods99c}.
Moreover, it does not show the high-frequency 
(10--1000 Hz) erratic variability which is detected from GF98, although 
this might be due to lower counting statistics.
However the two flares show also similar spectral properties. The photon spectrum of IF01
and the corresponding spectrum of GF98 during the first 40~s (corresponding
to the time duration of IF01) are both best fit with a three--component
({\sc bb + bknpl + pl}) model, with similar {\sc bb} temperature ($\sim 15$ keV)
and break energy ($E_{break} \sim 73$~keV) of the {\sc bknpl} model. However,
the power-law indices of the {\sc bknpl} are different, resulting higher (and the 
spectrum softer) for IF01. The highest energy sections of the spectra ($>300$ keV)
are both well fit with a power-law with marginally similar positive photon
indices. This power-law component contributes, in both cases, to $\sim 2$\% of the 
40--700 keV fluence of the two flares. 

In the magnetar model scenario, the entire 2001 flare and the intermediate
stage of GF98 (before the pulsation is set up clearly) could
be both interpreted as radiation coming from a transient pair-dominated corona
surrounding a trapped fireball anchored to the neutron star surface,
although other interpretations can be possible.
The high-frequency noise, detected during the intermediate stage of
GF98, could directly trace the evolution of fractures propagating
throughout the neutron star crust soon after the dramatic spike.
The PSD of the non--Poissonian noise is in agreement with the expectations
of the ``Integrate And Fire'' (IAF) model \cite{Davidsen02}, according to which,
similarly to earthquakes, discrete energy releases occur when a variable
threshold is exceeded.

The appearance of a last isolated pulse at the end of IF01 might point to
a different origin from what caused the flare: its peculiar time profile
is similar to that of short bursts from the same source.
While the main time profile of IF01 could have come from the outer layer
of the pair corona, the last pulse might have originated close to the surface.

\acknowledgements
We thank Sandro Mereghetti for carefully reading this manuscript and for his
comments.
This research is supported by the Italian Space Agency (ASI) and
Ministry of University and Scientific Research of Italy.
We wish to thank the Mission Director L. Salotti and the teams of the \sax\
Operation Control Center, Science Operation Center and Scientific Data
Center for their support.

%%%%%%%%%%%%%%%%%%%%%%%%%%%%%%%%%%%%%%%%%%%%%%%%
% BIBLIOGRAPHY
%%%%%%%%%%%%%%%%%%%%%%%%%%%%%%%%%%%%%%%%%%%%%%%%

\end{document}